\documentclass[11pt,a4paper]{article}

\usepackage{geometry} 
\usepackage{amsmath,amsfonts,amsthm,amssymb, mathtools, mathrsfs} 
\usepackage[dvipsnames]{color, xcolor}  
\usepackage{enumitem} 
\usepackage{graphicx, float, subfig, overpic} 
\usepackage{titlesec} 
\usepackage[none]{hyphenat} 
\usepackage[font=small,labelfont=bf,tableposition=top]{caption}
\usepackage{todonotes, comment} 
\usepackage{enumitem}

\usepackage{hyperref} 
\hypersetup{
	colorlinks=true,
	linkcolor=blue,
	citecolor=green,
}

\def\equal{:=}

\def\cO{{\mathcal O}}

\def\R{\mathbb R}
\def\Z{\mathbb Z}
\def\C{\mathbb C}

\def\im{{\rm i}}
\def\hp{\hat p}
\def\hq{\hat q}

\def\om#1{\omega\left(\frac{#1}{N}\right)}

\def\spazio#1{\ell^{#1}}
\def\reg{{\sigma}}

\definecolor{beatrice}{rgb}{1.0, 0.0, 0.5}
\definecolor{santiago}{rgb}{0.0, 0.0, 1.0}
\numberwithin{equation}{section}
\theoremstyle{plain}
\newtheorem{lemma}{Lemma}[section]
\newtheorem{corollary}[lemma]{Corollary}

\newtheorem{theorem}[lemma]{Theorem}

\newcommand{\norm}[1]{\left\lVert#1\right\rVert}

\renewcommand{\Re}{\mathrm{Re\, }}

\definecolor{myGreen}{RGB}{0, 200, 0}
\definecolor{myOrange}{RGB}{255, 100, 0}
\definecolor{myYellow}{RGB}{255, 200, 0}
\definecolor{myBlue}{RGB}{0, 200, 255}
\definecolor{myPurple}{RGB}{126, 47, 142}

\newcommand{\di}{\mathrm{d}}
\newcommand{\veps}{\epsilon}

\title{A survey on rigorous results for the dynamics of periodic FPU
chains} 
\author{Dario Bambusi\footnote{Dipartimento di Matematica,
  Universit\`a degli Studi di Milano, Via Saldini 50, I-20133 
  Milano. { \tt{email:dario.bambusi@unimi.it}}}
  \and
  Andrea Carati\footnote{Dipartimento di Matematica,
  Universit\`a degli Studi di Milano, Via Saldini 50, I-20133 
  Milano. { \tt{email:andrea.carati@unimi.it}}}
  \and
  Alberto Maiocchi\footnote{ Dipartimento di Matematica e Applicazioni
  - Edificio U5, Università degli Studi di Milano-Bicocca, via Roberto
  Cozzi, 55, 20125, Milan,
  Italy. \texttt{email:alberto.maiocchi@unimib.it}
  }
}
\date{\today}

\begin{document}

\maketitle
\begin{abstract}
In this paper we review some analytic results on the dynamics of the
FPU system. In the first part of the paper, having in mind that the
FPU Hamiltonian and the Toda Hamiltonian are close each other, we present some
results on the action angle variables of the Toda system and deduce
some stability properties for the dynamics of the FPU system. We first
focus on the case of finitely many particles and then we study the
limit $N\to\infty$. We present also some results on the continous
limit of the Toda chain showing that it is well described by a couple
of KdV equations. Then we study directly the dynamics of the function
interpolating the FPU system and show that the dynamics is Hamiltonian
and that the Hamiltonian is very close to a function of the first
three Hamiltonians of the KdV hierarchy. In the second part of the
paper we present some results valid in the thermodynamic limit,
according to which the time autocorrelation functions of some suitably
constructed observables decay slowly implying lower bounds on the
thermalization times of the system.
\end{abstract}


\section{Introduction}\label{intro}

In this paper we review some recent results on the dynamics of the
Fermi Pasta Ulam model. There is a huge number of papers studying the
dynamics of the FPU model mostly with the aim of understanding in a
heuristic or rigorous way the result of the computer simulation of
\cite{FPU} according to
which a chain of coupled particles attains a long lived metastable
state and only over a different time scale possibly attains
thermalization. The computer simulation in \cite{FPU} was done with 64
particles, but of course the main problem is that of the
persistence of the phenomenon in the thermodynamic limit in which the
number of particles $N$ tends to infinity and at the same time the 
total energy of the system diverges proportionally to $N$. 

Up to now there are no rigorous results which persist in the
thermodynamic limit and explain the FPU simulations, but some
phenomena have been understood mathematically and several dynamical
results are known. In the present paper we will present two kind of
results: the first dealing with the dynamics of the FPU at finite small energy,
and the second dealing with the problem of the fluctuations and the
decay of correlations at equilibrium.

\vskip 10 pt

The first group of result is mainly related to the approximation of
the FPU system with the Toda system and with the KdV equation.

We will first work at fixed number of particles $N$ and present some
results on the existence of action angle variables for the Toda
system. Using such variables one can apply KAM and Nekhoroshev theory
considering FPU as a perturbation of Toda. We will give a precise
statement of the Nekhoroshev's theorem one deduces for the FPU system.

Then we will discuss the behavior of these results when the number of
particles tends to infinity. In particular we will present a result on
the behavior of the action angle coordinates for the Toda system when
$N\to\infty$. It turns out that they develop a singularity at a distance of order
$N^{-3/2}$ from the origin. We will also deduce some results for the
FPU dynamics.

Then we will study the related problem of the continuous
interpolation. We will first consider the interpolation of the Toda
lattice and present some results showing that the actions and the
frequencies of the Toda system converge in the continuum limit to the
actions and the frequencies of a couple of KdV equations. These
results hold in the regime where the total energy of the system is of
order $N^{-2}$.

Then we will study the continuous limit directly on the FPU system and
present some recent results according to which the FPU dynamics is
well approximate up order $6$ in a suitable small parameter by a
combination of the first three equations of the KdV hierarchy.

Finally we present some nonrigorous results on the Fourier spectrum of
the FPU system which have been recently obtained exploiting the fact
that in some limit the FPU is described by the Burgers equation and
some recent results on the deduction of the wave kinetik equation for
the $\beta$ FPU system.

\vskip 10 pt

The second group of results contains a set of results on the stability for long
times of the orbits of the FPU system and related models in a
statistical sense, and is presented in Sect.~\ref{termo}. The focus is
set on time auto correlation of a dynamical variable at 
thermodynamic equilibrium, and the failure of
the autocorrelation of some physically relevant variables to decay to
zero in a given macroscopic time is seen as an indicator that, up to
that time, thermalization cannot be attained.

We will discuss the results obtained so far on the slow decay of
correlations for the energy of suitable ``packets'' of modes for the FPU system
and for the FPU system with alternating masses, for which the times
are much longer if the ratio of the heavy and the light mass is
large. Besides these results, valid for long, but finite times, we
present a related analytical method which allows to control also the
asymptotic decay of correlations, and has been used to study
numerically such a
decay in the FPU system. 

\section{Chains of particles with first neighbor interaction}\label{sect:FPU}

Consider the Hamiltonian of the FPU chain of $N$ particles with
periodic boundary conditions, namely
\begin{equation}
  \label{ham:FPU}
H(p,q):=\sum_{j=1}^{N}\frac{p_j^2}{2}+\sum_{j=1}^NV(q_{j+1}-q_j)\ ,
\end{equation}
where 
$q_{j+N}=q_{j}$, $p_{j+N}=p_{j}$ and $V$ is a potential.
Then the Hamilton
equations are
\begin{equation}
  \label{eq:FPU}
\ddot q_j=V'(q_{j+1}-q_j)-V'(q_j-q_{j-1})\ ,\quad j=1,...,N\ .
\end{equation}

In the case of the original FPU model one has
\begin{equation}
  \label{V}
V(\delta)=V_{FPU}(\delta):=\frac{\delta^2}{2}+\frac{\delta^3}{6}+\beta\frac{\delta^4}{4!}\ .
\end{equation}
In this case we will write $H=H_{FPU}$. We decompose
the FPU Hamiltonian as follows
\begin{equation}
\label{H}
H_{FPU}=H_0+H_1+H_2
\end{equation}
where
\begin{eqnarray*}
H_0&\equal &\sum_{j}\left(\frac{p_j^2}2+\frac{\left(q_{j+1}-
  q_j\right)^2}2
\right)\ ,\\
H_1&\equal& \frac 1{3!}\sum_{j}\left(
q_{j+1}- q_j\right)^3\\
H_2&\equal& \frac \beta{4!}\sum_{j} \left( q_{j+1}- q_j\right)^4\ .
\end{eqnarray*}

Introduce the discrete
Fourier transform defined by
\begin{equation}
\label{fou}
\hat{q}_k=\frac{1}{\sqrt{N}} \sum_{j=0}^{N-1}{q_j e^{2 \im \pi jk/N}},
\qquad k\in\Z\ ,
\end{equation}
and consider $\hat p_k$ defined analogously.  One
has $\hp_k = \hp_{k+N}, \, \hq_k
= \hq_{k+N}$, $\forall k \in \Z$, so we restrict to $\{\hat p_k, \hat
q_k \}_{k=0}^{N-1}$. Corresponding to real sequences $(p_j,q_j)$ one
has $\overline{\hat{q}_k}=\hat{q}_{N-k}$ and $
\overline{\hat{p}_k}=\hat{p}_{N-k}$.

Introduce the real Fourier variables  
\begin{equation}
\label{lin.bir.real}
X_k=\frac{\hat p_{k}+\hat p_{N-k}-\im\omega_k(\hat q_k-\hat q_{N-k})
}{\sqrt{2\omega_k} }\ ,\quad  Y_k=\frac{\hat p_{k}-\hat
  p_{N-k}+\im\omega_k(\hat q_k+\hat q_{N-k}) 
}{\im \sqrt{2\omega_k} }\ ,\quad k=1,...,N-1\ ,
\end{equation}
where $\omega_k\equiv\om k:=2\sin(k\pi/N)$; using such coordinates,
which are symplectic, the quadratic part
\begin{equation}
\label{h0}
H_0:=\sum_{j=0}^{N-1}\frac{p_j^2+(q_{j+1}-q_j)^2}{2}
\end{equation} 
of the Hamiltonian takes the form
\begin{equation}
\label{quad.part.0}
H_0
= \sum_{k=1}^{N-1}\omega_ k \frac{X_k^2+Y_k^2}{2}+\frac{\hat p_0^2}{2}\ .
\end{equation}

We remark that due to translation invariance the quantity 
\begin{equation}
  \label{P}
\sum_{j}p_j=P=\hat p_0 \sqrt{N}
  \end{equation}
is an integral of motion and as a consequence the ``center of mass
\begin{equation}
\label{media}
\sum_jq_j=Q=\hat q_0 \sqrt{N}\ ,\quad 
\end{equation}
moves uniformly. We restrict here to the case $P=0$, so also $Q=0$ is
invariant, so that the Hamiltonian \eqref{quad.part.0} takes the form
\begin{equation}
\label{quad.part}
H_0
= \sum_{k=1}^{N-1}\omega_ k \frac{X_k^2+Y_k^2}{2}= \sum_{k=1}^{N-1}E_k\ ,
\end{equation}
where
$$
E_k:=\omega_ k \frac{X_k^2+Y_k^2}{2}
$$


\section{FPU and Toda}\label{toda}

In the case where the potential in \eqref{ham:FPU} is exponential,
namely
\begin{equation}
  \label{vtoda}
V(\delta)=V_{Toda}(\delta):=e^{-\delta}
\end{equation}
one gets the Toda lattice, which is well known to be integrable. In
this case we shall denote the Hamiltonian by $H_{Toda}$.

The
trivial remark that
$$ \left|V_{Toda}(\delta)-V_{FPU}(\delta)\right|\leq
(1-\beta)\delta^4+ \cO(\delta^5)
$$
gives the connection with the FPU system and allows to use results on
Toda in order to deduce results on the FPU system. Actually in the
following we always concentrate on the case where $\beta\not=1$. When
$\beta$ approaches 1 one can actually improve the results.

\subsection{Finite number of particles $N$.}  The first idea is
to use KAM and the Nekhoroshev theory considering FPU as a
perturbation of the Toda lattice. To this end one has to know that the
Toda Hamiltonian, when written in action angle variables, fulfills the
nondegeneracy properties required for the application of the KAM or the
Nekhoroshev theorems. A proof of the fact that FPU fulfills the KAM
nondegeneracy property was done by Rink \cite{Rin01,Rin03} exploiting the
symmetries of the FPU system, while a detailed study of the action
angle variables for Toda, as well as a proof that they are
globally defined was done by Henrici and Kappeler. We state here the
result that was proved in \cite{HK08}.

\begin{theorem}[{{\cite{HK08}}}]
\label{HK} For any integer $N \geq 2$ there exists a global real analytic symplectic diffeomorphism
$\Phi_N:\R^{N-1}\times\R^{N-1} \to \R^{N-1}\times\R^{N-1}$, $(X,Y)=\Phi_N(x,y)$
with the following properties:
\begin{itemize}
\item[(i)] The Hamiltonian
  $$\tilde H_{Toda}:=H_{Toda}\circ \Phi_N
  $$ is a function of the
  actions $I_k^{Toda}:=\frac{x^2_k+y_k^2}{2}$ only.
\item[(ii)] The differential of $\Phi_N$ at the origin is the identity.
 \end{itemize}
\end{theorem}

Actually Henrici and Kappeler also computed  explicitly the
quadratic term of the Taylor expansion of $\tilde H(I^{Toda})$ at the
origin. Such an expression is different in the case of an even number
of particle \cite{HK09} and in the case of an odd number of
particles \cite{HK08a}, but in both cases the authors succeeded to
prove that such a Hamiltonian is convex as a function of the action
variables \cite{HK09a}. In particular one can thus apply both KAM
and Nekhoroshev theorem, still restricting to the manifold
$Q=P=0$. We state just the corresponding Nekhoroshev theorem. The
result is due to Henrici Kappeler \cite{HK09a} and the stability
exponent that we give are computed applying the version of
Nekhoroshev theorem due to Lochak and Neishtadt \cite{CN92} (see also
\cite{GCB}).

\begin{theorem}
  \label{Nek}
Consider the Cauchy problem for the FPU system \eqref{ham:FPU} with $\beta\not=1$, then
there exist positive constants $\epsilon_*$, $C$, such that for  any initial
datum $(p^0,q^0)$ fulfilling $P=Q=0$ and also 
\begin{equation}
  \label{soglia}
\epsilon:=\left\|(p^0,q^0)\right\|<\epsilon_*\ ,
\end{equation}
one has
\begin{equation}
  \label{sti.nek}
\left|I_j^{Toda}(t)-I^{Toda}_j(0)\right|\leq C\epsilon^{2+b}\ ,\quad \forall |t|\leq exp\left(\frac{\epsilon_*}{\epsilon}\right)^a
\end{equation}
with $I^{Toda}_j$ the actions of the Toda system and
\begin{equation}
  \label{esponenti}
a=\frac{1}{2(N-1)}=b\ .
  \end{equation}
\end{theorem}
We remark that a slightly better exponent $a$ was obtained in
\cite{ZZ17} at the expenses of worsening $b$. In any case one always has
\begin{equation}
  \label{a}
a\sim \frac{1}{2N}\ .
  \end{equation}

Now, the main interest when studying the FPU chain pertains the limit
$N\to\infty$, and the question is whether the above results, namely
Theorem \ref{HK} and \ref{Nek}, persist when $N\to\infty$.

It is very easy to see that Theorem \ref{Nek} becomes empty when
$n\to\infty$, indeed both exponents $a$ and $b$ tend to zero and thus
both the control of the actions (which are of order $\epsilon^2$) and
the time of control, become empty.

However the problem is even harder, indeed also Theorem \ref{HK} gives
a result for the case of a fixed number of particles, so it is
not at all clear if it remains valid in the limit $N\to\infty$.

\subsection{Limit $N\to\infty$.}\label{limit}
The behavior of the transformation introducing Birkhoff coordinates
for the Toda system in the limit $N\to\infty$ has been studied in
\cite{BM16}. It was proved that, on the one hand it develops a
singularity at a distance of order $N^{-2}$ from the origin, but on the
other hand it is well defined in a ball of such an order of magnitude,
even in a very strong topology and this allows to deduce a result on
FPU, which, up to now is the best available concerning the times over
which the dynamics is controlled. To state such results in a slightly
simplified, but precise way, for any $\sigma\geq0$, we introduce in
$\C^{N-1}\times \C^{N-1}$ the discrete Sobolev-analytic norm
\begin{equation}
\norm{(\hat p,\hat q)}^2_{\spazio{\reg}}:=\frac{1}{N}\sum_{k
  \not =0}[k]_{N}^{2}\, e^{2\sigma |k|}\omega_k\, \frac{|X_k|^2+|Y_k|^2}{2}
\label{nor.bir}
\end{equation}
where
$$
[k]_{N}:=\min\left\{k,N-k\right\}\ .
$$
The space $\C^{N-1}\times \C^{N-1}$ endowed
by such a norm will be denoted by $\spazio{\reg}$. 
We denote 
by $B^{{\reg}}(R)$  the ball of
radius $R$ and center $0$ in the topology defined by the norm
$\norm{.}_{\spazio{\reg}}$.

From now on we will denote
\begin{equation}
  \label{mu}
\mu:=\frac{1}{N}\ ,
\end{equation}
which will play the role of a small parameter. 

\begin{theorem}
\label{BMToda}
For any $\sigma\geq 0$ there exist strictly positive constants
$R_{*}$, $C$, such that for any $N\geq 2$, the map $\Phi_N$ of
Theorem \ref{HK} is analytic as a map from $B^{\reg}(R_{*}\mu^{3/2})$ to
$\spazio{\reg}$ and fulfills
\begin{align}
\label{Phi_N.est}
\sup_{\norm{(x,y)}_{\spazio{\reg}} \leq R\mu^{3/2}
}{\norm{\Phi_N(x,y)-(x,y)}_{\spazio{ \sigma}}}  \leq
C {R^2\mu^{3/2}}\ ,\quad \forall R<R_{\reg} .  
\end{align}
The same estimate is fulfilled by the inverse map $\Phi_N^{-1}$.
\end{theorem}

Thus we have that, at least in a ball of radius of order $\mu^{3/2}$
the Birkhoff coordinates of the Toda system are well defined. We
remark that in such a ball the energy is of order $N^{-2}$.

One would like
to apply also Nekhoroshev's theorem, but here one has to face the
problem that the best known threshold for the applicability of such a
theorem decrease more than exponentially with the number of degrees of
freedom.

What one can easily prove is a result which exploits the fact that FPU is
a perturbation of Toda lattice with a perturbation of fourth
order. The following theorem has been proved in \cite{BM16} as a
corollary of Theorem \ref{BMToda}.

\begin{theorem}
  \label{FPU.BM16}
Fix $\sigma\geq 0$ and consider the Cauchy problem for the FPU system
with an initial datum $(p^0,q^0)$ and let $R$ be s.t.
$\left\|(p^0,q^0)\right\|_{\spazio{\reg}}=R\mu^{3/2}$. There
exists $R_*>0$, $T>0$ such that, if $R<R_*$, then
\begin{equation}
  \label{FPU.fin}
\left\|(p(t),q(t))\right\|_{\spazio{\reg}}\leq 2 R_*\mu^{3/2}  \ ,\quad \forall
\left|t\right|\leq \frac{T}{R^2\mu^4}\ .
\end{equation}
\end{theorem}
We remark that since the norm has an exponential weight, it follows
that the energy in the $k-th$ mode decreases exponentially with $|k|$
for the times \eqref{FPU.fin}. This is the phenomenon usually referred
to as existence of a metastable packet of modes. Actually the times of
order $\mu^{-4}$ obtained by this theorem represent the longest lower
estimate that has been rigorously proven for the time of existence of
the FPU metastable packet. 

\subsection{Toda and KdV}\label{TodatoKdv} It is known since a long
time that the KdV equation is obtained in some formal continuous limit
from  FPU and also from Toda. Since both Toda and KdV are
integrable, it is natural to investigate the possible convergence of
the integrable structure of Toda to the integrable structure of
KdV. This was done in \cite{BKP0,BKP1,BKP2}. In such papers it was
shown that in the continuous limit both the actions and the frequencies
of the Toda system converge to the actions and the frequencies of a
couple of KdV equations. We are now going to state this result.

First of all we interpolate the state of the Toda chain. It is
convenient to proceed as follows: given two sequences $q_j$, $p_j$, we consider two
functions $u(x)$ and $w(x)$ periodic of period 1 in $x$ with the
property that
$$
\delta_j:=q_{j+1}-q_j=\mu^2u(\mu j)\ ,\quad  p_j(x)=\mu^2w(\mu j)\ ,
$$
and define the functions
\begin{equation}
  \label{onde}
r=\frac{u+w}{\sqrt2}\ ,\quad s=\frac{u-w}{\sqrt2} \ ,
\end{equation}
then, as $\mu\equiv N^{-1}\to0$ it turns out that $r$ and $s$ converge to
solutions of the KdV equations, one controlling right going waves, the
other controlling left going waves. 

To state the corresponding result we denote again by $I_j^{Toda}$ the action
variables of the Toda system and we remark that associated to a system
of action angle coordinates one has an important dynamical object
which is the set of the frequencies, which are just the derivatives of
the Hamiltonian with respect to the actions and that control the
evolution of the corresponding angles. We denote by
$\omega^{Toda}_j=\omega^{Toda}_j(I^{Toda})$ the $j$-th frequency of the Toda system.

One has corresponding objects in the KdV hierarchy. Namely given a
periodic function  $u$ one can represent it in action angle variables,
which are common to all the equations of the KdV hierarchy. We denote
by $I_{j}^{KdV}=I_{j}^{KdV}(r)$ the $j^{th}$ action of a function $r=r(x)$
according to the KdV hierarchy. 

We have the following theorem

\begin{theorem}
  \label{thm.BKP}
Assume that $u$ and $w$ are of class $C^2$, then, provided $N$ is
large enough one has the following asymptotic
\begin{align}
  \label{asym.BKP.1}
I^{Toda}_j=\frac{4}{N^2}\left[I^{kdV}_j(s)+\cO\left(\frac{1}{N^{1/16}}\right)\right]\ ,\quad
0<j< N^{1/16}\ ,
\\
\label{asym.BKP.2}
I_j^{Toda}=\frac{4}{N^2}\left[I^{kdV}_j(r)+\cO\left(\frac{1}{N^{1/16}}\right)\right]\ ,\quad
N- N^{1/16}<j<N\ ,
\\
\label{asym.BKP.3}
I_j^{Toda}=\cO\left(\frac{1}{N^{5/2}}\right)\ ,\quad N^{1/16}\leq j\leq N- N^{1/16}\ .
\end{align}
\end{theorem}
Such a theorem shows that the actions with small index behave in the
limit as the actions of a corresponding KdV equation. The
remaining actions tend to zero in the limit faster than $N^{-2}$. 

Concerning the dynamics, we consider the frequencies of the Toda
lattice. They will be approximated by the frequencies of a linear
combination of the equations of the first two equations of the KdV
hierarchy, so we recall them
\begin{align}
  \label{kdv1}
  K_1(u)&=\frac{1}{2}\int_0^1u^2dx\ ,
  \\
  \label{kdv3}
  K_2(u)&=\int_0^1\left(-\frac{1}{12}u^2_x+\frac{1}{3}u^3
  \right)dx\ ,
  \end{align}
and, given $c_2$, consider
\begin{equation}
  \label{vera}
K^{KdV}:=\frac{1}{N}   K_{1}+\frac{c_2}{N^3}
K_{2}\ .
  \end{equation}
 Consider the frequencies of such a Hamiltonian that will be denoted
by $\omega^{KdV}(u)$ (even if they depend only on the action
variables of $u$).  We also remark that since the KdV Hamiltonian are
scaled by a factor proportional to $N$, the frequencies are of order
$\mu$.

Concerning the frequencies we have the following result

\begin{theorem}
  \label{BKP.2}
There exists a value $c_2$ s.t., if $u$ and $w$ are $C^2$, then the frequencies of the Toda
system have the following asymptotic.
\begin{align}
  \label{asy.fre}
\omega_j^{Toda}&=\omega_j^{KdV}(s)+\cO\left(\frac{1}{N^{1/16}}\right)\ ,\quad
0<j<N^{1/64}\ ,
\\
  \label{asy.fre.1}
  \omega_j^{Toda}&=-\omega_j^{KdV}(r)+\cO\left(\frac{1}{N^{1/16}}\right)\ ,\quad
 N -N^{1/64}<j<N\ ,
\\
\label{asy.fre.2}
\omega_j^{Toda}&=\frac{2\pi j}{N}+\cO\left(\frac{n^3}{N^3}\right)\ ,\quad
N^{1/8}<j<N^{1/64}\ \text{and}\quad N-N^{1/64}<j<N-N^{1/8} 
\\
\label{asy.fre.3}
\omega^{Toda}_j&=2\sin\left(\frac{\pi
  j}{N}\right)\left(1+\cO\left(\frac{\log N^{1/16}}{N^{1/8}}\right)\right) \ ,\quad
N^{1/64}<j<N-N^{1/64}\ .
\end{align}
\end{theorem}
We remark that the choice of the powers $1/16$ or analogous quantities
has been done here in order to simplify the statements: in the original
statement one has quite a lot of freedom to make different
choices.

\section{Continuous limit and KdV}\label{inter}

The results of the previous section show that in some precise sense
Toda converges to KdV in the continuum limit and thus the same is
true for the FPU. In this section we analyze
how the equations of the KdV hierarchy can be obtained directly
working on the FPU system. Quite surprisingly it turns out that the
order of approximation can be pushed further.

\subsection{Interpolation}
First we rewrite the FPU equations in terms of the variables
\begin{equation}
  \label{rj}
\delta_j:=q_{j+1}-q_j\ ,
\end{equation}
namely
\begin{equation}
  \label{ddotr}
\ddot \delta=\Delta_1V'(\delta)\ ,
\end{equation}
with $\Delta_1$ the operator of finite difference, namely
$$
(\Delta_1 \delta )_j:=\delta_{j+1}+\delta_{j-1}-2\delta_j\ .
$$
Then we interpolate and scale amplitude and time: let $
u(\tau,x)$ be a smooth function periodic of period $1$ in $x$ with the property that
\begin{equation}
  \label{interpolo}
\delta_j(t)=\epsilon u(\mu t, \mu j)\ ,
\end{equation}
with $\mu:=N^{-1}$. 
In order to get that $\delta$ fulfills the correct equations we postulate
for $u$ the equations
\begin{equation}
  \label{interpolo.1}
u_{\tau\tau}=\Delta_\mu \frac{V'(\epsilon u( x))}{\epsilon}\ ,
\end{equation}
where $\Delta_\mu$ is the finite discretization of the second
derivative, namely
$$
(\Delta_\mu u)(x):=\frac{u(x+\mu)-2u(x)+u(x-\mu)}{\mu^2}\ .
$$
It is easy to see that this equation is
Hamiltonian with Hamiltonian function
\begin{equation}
  \label{Hamiltoniana}
\int_0^1\left\{ \frac{v(x)[-\Delta_\mu v(x)]}{2}+\frac{V(\epsilon u(x))}{\epsilon^2}\right\}dx\ ,
\end{equation}
where $v$ is the variable conjugated to $u$.

We remark that if $\epsilon=\mu^2$ then one has $w=v_x+\cO(\mu)$,
where $w$ is the function interpolating the sequence $p_j$. 

Expanding in power of $\epsilon$ and $\mu$ we get
\begin{align}
  \label{}
H(u,v)=H_0(u)+\epsilon H_1(u)+\epsilon^2 H_2(u)+\text{higher\ order\ terms}\ ,
\end{align}
with
\begin{align}
  \label{H0}
  H_0(u,v)&=\int\frac{v_x^2+u^2}{2}dx\ ,
  \\
  \label{H1}
  H_1(u,v)&=\frac{1}{6}\int\left(\frac{\mu^2}{\epsilon}v_{xx}^2+u^3\right)dx\ ,
  \\
  \label{H2}
  H_2(u,v)&=\int \left(\frac{\mu^4}{\epsilon^2}\frac{v_{xxx}^2}{45}+\frac{\beta}{4!} u^4  \right)dx\ .
\end{align}
Following \cite{Craig,BP06} it is very useful to make the non canonical
transformation introducing new variables $(r(x),s(x))$ defined by
\begin{equation}
  \label{r,s}
r=\frac{u+v_x}{\sqrt2}\ ,\quad s=\frac{u-v_x}{\sqrt2} \ .
\end{equation}
We remark that these quantities do not coincide with the quantities
$(r,s) $ introduced in Subsect. \ref{TodatoKdv}: equality holds only up
to $\cO(\mu)$. 

In the variables $(r,s)$ the Poisson tensor transforms in such a way
that the Hamilton equations of a Hamiltonian $H$ become
\begin{equation}
  \label{gard}
\frac{dr}{d\tau}= -\partial_x\nabla_r H\ ,\quad\frac{ds}{d\tau}= \partial_x\nabla_s H\ ,
\end{equation}
where $\nabla_r$ is the $L^2$ gradient with respect to the variable $r$
and similarly $\nabla_s$.
The various part of the Hamiltonian take the form
\begin{align}
  \label{1Kdv}
H_0&=\int_0^1\frac{r^2+s^2}{2}dx
\\
  \label{2Kdv}
H_1&=\frac{1}{12}\int_0^1\left[\frac{\mu^2}{\epsilon}r_x^2+\frac{r^3}{\sqrt2}
  \right]dx
+
\frac{1}{12}\int_0^1\left[\frac{\mu^2}{\epsilon}s_x^2+\frac{s^3}{\sqrt2}\right]dx
\\
  \label{3Kdv}
  &+\int_0^1\left[\frac{1}{4\sqrt2}(r^2s+s^2r)-\frac{1}{6} \frac{\mu^2}{\epsilon}
    r_xs_x\right]dx
  \\
  \label{4Kdv}
  H_2&=\int_0^1\left[\frac{\mu^4}{\epsilon^2}\frac{(r_{xx}-s_{xx})^2}{90}+\beta\frac{(r+s)^4}{96}\right]dx \ .
\end{align}
We write explicitly the equation of $r$, the one for $s$ being
similar (one has to exchange $s$ with $r$ and add a minus sign): 
\begin{align}
\label{leduekdv}
r_{\tau}&=-r_x+\frac{\epsilon}{6}\left(\frac{\mu^2}{\epsilon}r_{xxx}-
\frac{3}{\sqrt2}rr_x\right)-\epsilon^2 \left(\frac{\mu^4}{\epsilon^2}\frac{\partial^5_xr}{45}+\frac{\beta}{24}\partial_xr^3\right)
\\
  \label{altro}
&-\frac{\epsilon}{4\sqrt2}(2rs+s^2)-\frac{\mu^2}{6}s_{xxx}+\frac{\mu^4}{45}s_{xxxxx}-\frac{\epsilon^2\beta}{96}\partial_x
  \left[12r^2s+12rs^2+4s^3  \right]\ .
\end{align}
First we remark that if we neglect the coupling \eqref{altro}, then
one gets an equation for $r$ alone, furthermore such an equation is a
perturbation of order $\epsilon^2$ of a KdV equation in a moving frame. A
naive idea to analyze the dynamics (but which in any case leads to
interesting results) consists in assuming that $s$ and its derivatives
are very small for all the times we are looking at. Then, passing to a
moving frame, namely defining $y=x-\tau$, one is led to the equation 
\begin{equation}
  \label{una.sola}
\frac{6}{\epsilon}r_{\tau}=\frac{\mu^2}{\epsilon}r_{yyy}-
\frac{3}{2\sqrt2}rr_y\ ,
\end{equation}
which, after rescaling again time, is a KdV equation in the zero
dispersion limit $\mu\to 0$. This is a quite complicated problem to
which we will come back in Sect. \ref{GPR}.

\subsection{$KdV_3$: a good approximation}\label{KdV5}

In the case $\epsilon=\mu^2$ the limiting KdV equation is no more
singular. Furthermore it has been shown in a series of paper that one
can actually get rid of the coupling terms between $r$ and $s$ by a
procedure of Birkhoff normal form. Such a procedure can be carried out
until the order $\mu^4$, while afterwards one gets some
obstructions. Furthermore one can use Kodama theory \cite{Kodama} in
order to show that, after a change of coordinates the FPU equations are
very well approximated by a function of the first three equations of
the KdV hierarchy. In order to state precisely the corresponding
result we start by recalling that the first three Hamiltonians of the
KdV Hierarchy are \eqref{kdv1}, \eqref{kdv3} and
\begin{align}
  \label{kdv4}
  K_3(u)&=\int_0^1\left(\frac{1}{2}u^2_{xx}-\frac{5}{2}u^2_xu+\frac{5}{8}u^4
  \right)dx\ ,
\end{align}
and the corresponding equations are obtained by using Gardner
Hamiltonian structure, namely they are given by
\begin{equation}
  \label{eq.KdV}
u_\tau=\partial_x\nabla K_n\ . 
\end{equation}
One recognizes that for $n=0$ one gets the transport equation, while
for $n=1$ one gets the standard KdV equation and for $n=2$ one gets
the second equation of the KdV hierarchy. We recall that these
equations are commuting, in the sense that the function $K_n$ is an
integral of motion for each one of the equations of the Hierarchy and
also for the dynamics of any function of the Hamiltonians of the KdV
hierarchy.

The Hamiltonian that approximate the FPU is constructed as follows:
given constants $c_2,c_3,c_4,c_5$ define
\begin{equation}
  \label{QuasiQuellaGiusta}
K(r):=K_1(r)+\mu^2 c_2 K_2(r)+\mu^4 \left(c_3K_3(r)+c_4[K_2(r)]^2\right)
\end{equation}
and then consider
\begin{equation}
  \label{QuellaGiusta}
K(r)+K(s)+c_5\mu^4 [K_2(r)][K_2(s)]\ .
\end{equation}

In order to state the main theorem we use the Sobolev norms $H^s$
which are defined in the standard way. We will denote by $B_s(R)$ the
ball of radius $R$ centered at the origin with respect to the $H^s$
norm. The following theorem is due to Gallone, Ponno, Rink \cite{GPR21,Gallone2022}

\begin{theorem}
  \label{th:Kdv5}[Gallone, Ponno, Rink]
There exists $s'$ and a map
$T_\mu:B_{s+s'}(R)\times B_{s+s'}(R)\to B_s(R)\times B_s(R)$, with the
following properties
\begin{itemize}
\item[(i)] $\sup_{{(r,s)\in B_{s+s'}(1)}}\|{T_\mu(r,s)-(r,s)}\|_{H^s
}\leq C \mu^2$,
\item[(ii)] There exist constants $c_2,...,c_5$ with the following
  property: let $I$ be a possibly infinite interval of rescaled time containing the origin and $z(.)=(r(.),s(.))\in C^1(I;B_{s+s'}(1))$ be a solution
  of the Hamiltonian system \eqref{QuellaGiusta}
define
  \begin{equation}
  \label{trans}
z_a\equiv(r_a,s_a):=T_{\mu}(r,s)\ .
\end{equation}
Then there exists $R\in C^1(I,H^s\times H^s)$ s.t. one has
  \begin{equation}
    \label{main.diff}
\dot z_a(\tau)=J\nabla H_{FPU}(z_a(\tau)) + \mu^6 R(\tau)\ ,\quad \forall \tau\in
I\ ,
  \end{equation}
  where $H_{FPU}$ is the Hamiltonian \eqref{1Kdv}-\eqref{4Kdv} of the FPU system
  rewritten in the variables $(r,s)$.
\end{itemize}
\end{theorem}

A similar result for the water wave problem on $\R$ was proved in
\cite{Bam21}.

Then, following the proof of the main result of \cite{BP06} one can
probably deduce a corollary showing that the solution of the FPU
system remains $\cO(\mu^{3})$ close, to the solutions constructed with
the Hamiltonian \eqref{QuellaGiusta}, for a time of order $\mu^{-3}$. However, a big
limitation of such a result is that, notwithstanding the fact that the
error in the equations is of order $\mu^6$ the approximation, remains
valid only over a time scale $\mu^{-3}$, a time over which the effects
of the $KdV_3$ are at most of order $\mu$, so not yet visible at a macroscopic scale. This is a
limitation that the theory shares with all the known results relative
to the study of modulation equations.

A further  worst limitation is that the result is valid only in a regime of
energy going to zero as $N\to\infty$.

\section{Turbulent behavior and approach to equilibrium}\label{turbo}

\subsection{Continuous limit and approach to equilibrium}
\label{GPR}

As shown in \cite{GPRuffo} the continuous limit can be obtained also in
order to deduce some interesting results on the way to thermalization
of the FPU model. To present their result consider again the equation
\eqref{una.sola}: by rescaling the time to $\tilde \tau:=\epsilon\tau/6$
and the amplitudes, one is led to the rescaled form
\begin{equation}
  \label{una.sola.1}
r_{\tilde\tau}=\frac{\mu^2}{\epsilon} r_{xxx}-rr_x\ .
\end{equation}
The idea of \cite{GPRuffo} is to take the formal zero dispersion limit
$\mu=0$, which gives a Burgers equation, which should describe well
the solution of the FPU as far as the solution has a bounded third
derivative. It is well known that the solution of the Burgers equation
develops a singularity at finite time and this is the time at which,
for sure, the approximation fails. However the authors take advantage
of the fact that one can write explicitly the evolution of the
Fourier coefficients of solution of the Burgers equation and that the
corresponding expression does not have any singularity, thus one can
analyze the behavior of the Fourier coefficients also when the
solution in physical space is no more defined. In \cite{GPRuffo} the
authors compute explicitly the Fourier spectrum of the solution at the
time where the singularity in physical space appear and they find the
very surprising result that the energy in the $k$--th mode decreases
as $|k|^{-8/3}$, namely there is a power law decay, as predicted by
Kolmogorov theory of turbulence for the Navier--Stokes equation. They
also compare the result with numerical simulations and they find that
they are in a quite good agreement with the theoretical
predictions. The numerical simulations also give more results, but we
will not discuss them here.

We remark that it is not clear how to deduce a theorem from the above
method, but similar ideas have been used by Kuksin in order to discuss
the development of turbulence in NLS \cite{Kuk96}. Such ideas can be
extended to the present framework, in particular considering equation
\eqref{una.sola.1} as a perturbation of the Burgers' equation one can
prove the following theorem whose application to the FPU model is in
progress.

\begin{theorem}
  \label{Kuk}Consider the equation \eqref{una.sola.1}
  with initial datum $r_0(x)=\sin x$, and assume $\mu':=\mu^2/\epsilon$ small
  enough, then there exists a time $t_*$ such that the corresponding
  solution fulfills
  \begin{equation}
    \label{concludo.kuksin}
\max_{k\leq
  5}\sup_{x}|\partial^k_xr(x,t_*)|\geq\frac1{(\mu')^{1/4}}\ .
  \end{equation}
\end{theorem}
Such a theorem shows that, although in a very weak sense, one has a
transfer of energy to high frequency modes. The advantage is that the
result holds in the thermodynamic limit.

\subsection{Wave Kinetic Equation}\label{WKE} A very important result has
been obtained quite recently, namely the proof of the validity of the
Boltzmann equation for a gas of interacting particles \cite{DHM24}. The method is very robust
and it has also been used to deduce the validity of the wave kinetic
equation for the dynamics in the NLS equation \cite{DH23} and it has also been
applied recently to the dynamics of the $\beta$-FPU model
\cite{Wu}.

The idea of the deduction of the wave kinetic equation is to use a
variant of Birkhoff normal form to extract the relevant interactions
from the nonlinearity and then to make a random phase approximation in
order to extract the contributions which are relevant for almost all
the values of the phases. Indeed, it is well known that corresponding
to random choices of the phases most of the interactions among
different normal modes are negligible. Then the authors also take the
large limit box, which in the case of FPU is the limit
$N\to\infty$. In order to transform the procedure into a theorem the
most difficult part consists in proving that the random phase
approximation is preserved by the dynamics.

A detailed analysis of the FPU dynamics based on these ideas has been
done in a series of papers by Onorato et al. \cite{Ono15,Ono18} in which
the authors have been able to put into evidence the main interactions
driving the thermalization process in the FPU dynamics. We refer to
the original papers for more details.

As we anticipated, up to now not much is rigorously known on the
validity of the wave kinetic equation and the corresponding
thermalization processes in the FPU model. We just mention the result
by Wu \cite{Wu} ensuring the validity of the wave kinetic equation in
the $\beta$ FPU model, provided $\beta\sim N^{-\gamma}$ with some
positive $\gamma$. However this result is valid only up to a fraction
of the kinetic time, so that it does not guarantee that the predictions
of the WKE are rigorously justified in the dynamics of the FPU model.

We also mention the recent paper \cite{OnoHani} in which the
authors study the normal form transformation removing the nonresonat
interactions in the FPU system and they show that the
transformation they construct is well defined if $\beta\leq
N^{-(1+\epsilon)}$.

\section{Results at the thermodynamic limit}\label{termo}

The results of the previous Sections~\ref{toda} and \ref{inter} imply
that the dynamics of the system remains close to integrable 
for long times. In particular, it was shown that the solutions to the
equations of motion are well approximated by the solutions of a
suitable integrable system. The reported results hold in the limit in
which $N\to\infty$ with a total energy $E\sim N^{-a}$ with $a=2$ or
$3$. However the physically more relevant limit is the thermodynamic
limit, in which the specific energy $\veps=E/N$ is kept fixed as
$N\to\infty$. The study of this limit is what matters for statistical
mechanics, where, however, the object of study are no longer the
system's trajectories but its ergodic properties.

In the ergodic context, the equilibrium state is characterized by an
invariant measure $\mu$, and thermalization refers to the process by
which, starting from an initial measure $\mu_0$, the evolved one
$\mu_t$ converges to the invariant measure $\mu$. From this
perspective, the numerical experiment by Fermi-Pasta-Ulam shows that
starting from a measure concentrated just on the first few modes, as it
evolves, it does not converge to the microcanonical measure, which
weights all normal modes of the system equally, up to an extremely
long time scale. In this sense, this experiment suggests the
thermalization time to be extremely long at low specific energies.

To obtain rigorous results, one exploits the property that the
convergence of $\mu_t$ to $\mu$ is equivalent to the vanishing of the
time autocorrelation for any square-integrable
function. Therefore, by showing that the autocorrelation of a suitable
dynamic variable remains close to its initial value for long times, we
obtain an estimate from below of the thermalization time. The first
subsection reports the rigorous estimates obtained using this approach
for the FPU system in the thermodynamic limit.

But the study of autocorrelations has also a second aspect, in which
we report in the second subsection. In fact, according to linear
response theory, if a system starts from an equilibrium condition, the
response to an external force, or the way it adjusts to a new
equilibrium after a perturbation, are determined by the time correlations
of suitable dynamical variables, provided that the perturbation be
sufficiently small. For example, the electrical susceptibility of a
crystal is determined by the Fourier transform of the autocorrelations
of the optical normal modes\footnote{Optical normal modes are the
normal modes corresponding to that branch of the dispersion relation
that does not vanish at $k=0$. See below.}  $\hat{\mathbf{q}}_k^+$ with
$k=0$. In this way, the persistence of an integrable behavior of the
correlations over long times heavily influences the physically
observable quantities. In particular, the way in which the
autocorrelations vanish for $t\to\infty$, whether they decay
exponentially, rather than as stretched exponentials, or even as a
power law, is particularly relevant. Notice that the study of the
correlations decay is a very active field in experimental physics, for
what concerns for example, glasses, supercooled liquids and more in
general the field of soft matter (see for example the review
\cite{douglas}).

So, in the second subsection are reported some known analytical results
on the decay of time correlation, and in particular a criterion that ensures
that the decay is not exponential.

\subsection{Finite time stability of correlations}

The most straightforward example is the study of the thermalization time of
the system, that is, the time in which the system reaches
thermodynamic equilibrium after a dis\-pla\-ce\-ment from it. A lower bound
on this time can be obtained 
by providing an estimate of the time during which the time
autocorrelation of a dynamical variable remains far from zero. Recall that,
given a dynamical variable $f:\mathcal{F}\to \R$ and denoting by
$\Phi^t:\mathcal{F}\to\mathcal{F}$, the flow induced on the phase space
$\mathcal{F}$ by the solutions of the equations of motion, the
autocorrelation $\mathcal{C}_f(t)$ is simply defined as
\begin{equation*}
\mathcal{C}_f(t)=\big< f(t) f(0) \big> -\big< f \big>^2 \ ,
\end{equation*}
where we have denoted $f(t)\equal f\circ\Phi^t$, while the average
$\big < \cdot \big>$
is taken with respect to a flow-invariant $\mu$ measure. As always in
statistical mechanics, the invariant measure we will consider is the
Gibbs measure. As it is well known, it is defined starting from the
Hamiltonian of the system $H(q,p)$, using the
formula
\begin{equation*}
\di \mu = \frac{e^{-\beta H}}{Z(\beta)}\,\di^N q\di^N p\ ,
\end{equation*}
where the partition function $Z(\beta)$ is the normalization factor of
the measure (i.e. $Z=\int \exp(-\beta H)\di^N q\di^N p$) and the
parameter $\beta$ is usually identified with the inverse temperature
of the system.

An interesting result was obtained in \cite{bamcarmaio} concerning the FPU
systems: the energy of ``packets'' of normal mode (defined shortly)
remains correlated for times of the order of $\beta$, that is, the
thermalization time increases at least with the inverse of the
temperature as the temperature decreases towards zero. The idea of
considering a ``packet'' of normal modes as a significant dynamic
variable comes from the fact that in the limit $N\to\infty$ the
frequencies of the normal modes become a continuum, and therefore,
every measurement, due to the physical limitations, will always
concern a group of modes with approximately the same frequency.  Let
us then define the observable
\begin{equation*}
J_x =\sum_k g(k/N) E_k \ ,
\end{equation*}
where $g:[0,1]\to[0,1]$ is a smooth function defined by the interval
$[0,1]$ in itself, peaked around a value $x\in[0,1]$, while $E_k$ is
the energy of the $k$-th normal mode (see \eqref{quad.part}). $J_x$
therefore represents the (mean) action of the packet of oscillators of
frequency $\omega(x)$.

For this observable, the following Theorem holds (Corollary 1 of the
paper \cite{bamcarmaio}):
\begin{theorem}\label{t:bamcarmaio}
There exist constants $\beta^*>0$, $N^*>0$, and $C > 0$ such that, for
any $\beta> \beta^*$ and for any $N > N^*$, one has
\begin{equation*}
\mathcal{C}_{J_x}(t) > \frac 12 \sigma^2_{J_x} \quad \mbox{for
all}\ |t|\le C \beta \ .
\end{equation*}
\end{theorem}
Therefore, according to this theorem, the correlation time goes to
infinity as the temperature tends to zero. This estimate is probably
not optimal, because  numerical work in paper \cite{benchripon}
seems to indicate a more than linear growth of the correlation times
with $\beta$.

A similar result has been obtained in \cite{GMMP} by exploiting the
proximity between FPU and Toda potentials and considering the
time evolution of the integrals of motion for the Toda chain under the
flow induced by the closest FPU Hamiltonian. Denote by
$J^{(m)}$ the trace of the $m$--th power of the Lax matrix associated
to the Toda chain, divided by $m$ (see (2.12) in \cite{GMMP}). Such
quantities are integral of motions for the Toda Hamiltonian flow and have a
rather simple analytic expression in terms of the canonical variables
$(q_j,p_j)$. For them, an analog of Theorem~\ref{t:bamcarmaio} holds,
with $J^{(m)}$ replacing $J_x$ (see Theorem 2.1 in
\cite{GMMP}). Moreover, this provides in turn  an estimate involving
some special packets of normal modes, since $J^{(m)}$ is well
approximated, for low specific energies, by a suitable combination of
energies of normal modes. This is expressed by the following Theorem
(see Theorem 2.5 in \cite{GMMP}):
\begin{theorem}\label{t:bamcarmaio2}
Fix $m\in\mathbb N$ and let $\mathbf g=(g_{0},\ldots,g_{N-1})$ be a
vector such that $g_j=g_{N-j}$ for $0\le j\le m$, $g_j=0$ otherwise
and there exist $K,K'>0$ for which 
$K'<\sum_j|g_j| <K$. Consider the Hartley transform
$$
\hat g_k \equal \frac 1{\sqrt{m}}\sum_{j=-m}^m g_j\bigg( \cos
\Big(2\pi\frac {jk}{m}\Big) + \sin \Big(2\pi\frac {jk}{m}\Big) \bigg)
$$
of the vector $\mathbf g$  and define the function
$$
\Phi \equal \sum_k \hat g_k E_k\ .
$$
Then there exist constants $\beta^*>0$, $N^*>0$, and $C > 0$ such that, for
any $\beta> \beta^*$ and for any $N > N^*$, one has
\begin{equation*}
\mathcal{C}_{\Phi}(t) > \frac 12 \sigma^2_{\Phi} \quad \mbox{for
all}\ |t|\le C \beta \ .
\end{equation*}
\end{theorem}

The estimates of the previous theorems are obtained by exploiting just
the vicinity of the Toda and the FPU systems and this gives results
valid up to 4-th order corrections. Better estimates would be obtained
by carrying out the construction of adiabatic invariants at higher
orders. As is known, this involves small denominators which up to now
it is not know how to control. Nevertheless numerical simulations
seem to suggest that stability holds also at slightly higher order.

One model in which the small divisors problem can be overcome is the
FPU alternating-mass model, a model which goes back to Born and Von
K\'arm\'an \cite{BK} (see paper \cite{GGMV} for a recent study)
and studied in the thermodynamic limit in paper \cite{Maiocchi}.
One considers a model with two kind of particles with different
masses, whose Hamiltonian is the following
\begin{equation*} 
H=\sum_j \Big( \frac {p^2_{j,1}}{2m_1} + \frac {p^2_{j,2}}{2m_2}\Big) 
+ \sum_j V(q_{j,2}-q_{j,1}) + V(q_{j+1,1}-q_{j,2}) \ ,
\end{equation*}
with periodic boundary conditions, while the potential is assumed to
be quartic,
that is, $V(r)=r^2/2 + A r^3/3 + B r^4/4$. Normal modes can be introduced
to diagonalize the quadratic part $H_0$
of the Hamiltonian, which takes the form
\begin{equation*}
H_0 = \frac 12 \sum_k\left( \big|\hat p^+_k\big|^2 + \omega^2_{+}(k) \big|\hat q^+_k\big|^2\right) +
\frac 12 \sum_k\left( \big|\hat p^-_k\big|^2 + \omega^2_{-}(k) \big|\hat q^-_k\big|^2\right) \ ,
\end{equation*}
where $\hat p_k^\pm$ and $\hat q_k^\pm$, for
$k=\lfloor-N/2\rfloor+1,\dots,\lfloor N/2\rfloor$, are
canonically conjugate complex variables and the frequencies are given
by
\begin{equation*}
\omega_{\pm}(k) = \sqrt{
\frac {m_1+m_2\pm\sqrt{m_1^2+m_2^2+2m_1m_2\cos(2\pi k/N)}}{m_1m_2}
}
\end{equation*}

The branches $\omega_{-}(k)$
and $\omega_{+}(k)$, called acoustic and optical respectively,
are separate from each other, as the former
takes values between 0 and $\sqrt{2/m_1}$, while the latter
takes values greater than $\sqrt{2/m_2}$. Note that the distance
increases with the difference in masses, and it can also be seen that when
$m_1\gg m_2$ the frequencies of the optical band assume a
nearly constant value.

Therefore, for each integer $k$, there are two normal modes. Two
macroscopic dynamical variables can be introduced: the harmonic energy
$E^+$ of the ``optical" normal modes, defined by $E^+=1/2 \sum
\big|\hat p^+_k\big|^2 + \omega^2_{+}(k) \big|\hat q^+_k\big|^2$ (i.e.,
the normal modes corresponding to the frequencies of the optical
band), and the harmonic energy $E^-$ of the ``acoustic" normal modes,
defined by $E^-=1/2 \sum \big|\hat p^-_k\big|^2 + \omega^2_{-}(k)
\big|\hat q^-_k\big|^2$ (i.e., the normal modes corresponding to the
frequencies of the acoustic band). If the frequencies are very
different, the two subsystems are weakly coupled , and therefore their
energies change slowly over time.  In this way, they remain correlated
for a long time. An estimate of this time is given in the following
theorem (Theorem~1 of the paper \cite{Maiocchi})
\begin{theorem}
There exist constants $\beta^*>0$, $N^*>0$, $M>2$, and $C_1,C_2> 0$
such that, for any $\beta> \beta^*$, any $N > N^*$ and for any
$m_1/m_2>M$ one has
\begin{equation*}
\Big| \mathcal{C}_{E^\pm}(t) - \mathcal{C}_{E^\pm}(0) \Big|\le
C_1\Bigg( \frac 1{\sqrt{\beta}} + \frac {m_2}{m_1} \Bigg)
\sigma^2_{E^\pm}\quad \mbox{for}\ |t|\le C_2\beta^S \ ,
\end{equation*}
where $S=\lfloor\sqrt{m_1/m2}/2\rfloor$.
\end{theorem}
We therefore see that for sufficiently low temperatures, for a
sufficiently small mass ratio, the autocorrelations of the harmonic
energies $E^\pm$ are close to their initial value (and therefore
different from zero), for times that increase as a power of $\beta$:
the exponent becomes larger as the ratio between the two kind of
masses increase. This is due to the fact that the perturbative
construction can be carried to higher orders, unlike the case of equal
masses. In fact, in the case of alternating masses, the small divisors
appear at a perturbative order that is increasingly higher the greater
the distance between the acoustic and the optical band. And as already
mentioned, this distance depends on the mass ratio.

\subsection{Asymptotic decay of the correlations}
The results cited in the previous subsection, are based on the simple estimate introduced in
the paper \cite{Carati}
\begin{equation}\label{eq:deccor}
\mathcal C_f(t) \ge \sigma^2_f - \frac {t^2}2 \bigl\| [f;H] \bigr\|^2
\ ,
\end{equation}
where $[\cdot;\cdot]$ denotes the Poisson bracket, while, at the
variance with the notations used in the previous Sections, now
$\bigr\| \cdot \bigl\|$ denotes the norm $L^2$ in the phase space
$\mathcal F$ endowed with the invariant measure $\mu$. This estimate
can be thought of as an estimate from below of the slope of the curve
$\mathcal{C}_f(t)$. Although it allows us to estimate the
thermalization time, this relation does not allow us to control the
asymptotic nature of the decay, that is, the actual decay over long
timescales. In a certain sense, it only controls the initial stage of
the process of the decaying of correlation. The study of the
asymptotic nature of correlation decay, requires a generalization of
this relation. This generalization was given in the paper
\cite{cargiorgmaioc}, where the following theorem was proved. Defined
by recurrency the $k$-th order Lie derivative of $f$, for $k\ge 1$,
namely $f^{(k)}\equal[f^{(k-1)};H]$, where $f^{(0)}\equal f$, one has
\begin{theorem}\label{teor:serie}
Let $\mu$ be a probability measure on the phase space $\mathcal{F}$,
invariant for the flow generated by $H$, and let $n>0$. Then, for any
dynamical variable $f\in L^2(\mu, \mathcal{F})$ such that $f^{(k)}\in
L^2(\mu, \mathcal{F})$ for all $k\le n$, one has 
\begin{equation}\label{eq:serie_troncata}
\begin{split}
\mathcal C_f(t) = &\sigma^2_f
+\sum_{k=1}^{n}(-1)^k\bigl\|f^{(k)}\bigr\|^2\frac{t^{2k}}{(2k)!}\\ &+
(-1)^{n+1} \int_0^t\di t_1 \ldots \int_0^{t_{2n-1}} \di t_{2n} \bigl
\|f_{t_{2n}}^{(n)}-f^{(n)}\bigr\|^2\ .
\end{split}
\end{equation}
\end{theorem}

We therefore obtain an expansion with an integral remainder that has a
definite sign, negative for even $n$ and positive for odd $n$. This
allows to give bounds from above and below to the decay of the
correlations, once we know how to calculate (for instance,
numerically) the coefficients
\begin{equation}\label{eq:definitions_a_and_c}
c_0\equal \sigma_f^2 \ ,\quad c_n\equal \bigl\|f^{(n)}\bigr\|^2 \mbox{
for } n> 0\ .
\end{equation}

However, a more interesting theoretical use can be obtained from this
development, to establish the asymptotic character of the decay of the
correlations over long timescales, in particular, to decide whether
the decay is exponential or not. To this end, it is useful to consider
the Laplace transform $\hat{\mathcal C}_f(p)$ of the correlations
\begin{equation*}
\hat{\mathcal C}_f(p) \equal \int_0^\infty e^{-pt} \mathcal C_f(t)\di p
\ ,
\end{equation*}
since a necessary and sufficient condition for the decay to be
exponential is that the transform $\hat{\mathcal C}_f(p)$ be analytic
in a half-plane $\Re p >\gamma$ with $\gamma<0$. The connection
between the coefficients $c_n$ and the Laplace transform is
established in the following theorem (Theorem~3 of the paper \cite{cargiorgmaioc})
\begin{theorem}\label{teor:stieltjes}
  Let $\mathcal C_f(t)$ be analytic in $t$ about the origin and
  continuous for any $t\in \mathbb R$. Then the following statements
  hold:
  \begin{enumerate}\label{prop:stieltjes_1}
  \item in the half-plane $\operatorname{Re} p> 0$ the Laplace
    transform $\hat{\mathcal C}_f(p)$ is analytic and there exists a
    positive Borel measure $\alpha(s)$ such that 
    \begin{equation}\label{eq:stieltjes_laplace}
      \hat{\mathcal C}_f(p) = \frac {p}{\pi} \int_0^{+\infty} \frac
          {\di \alpha(s)}{p^2+s^2} \ ,
    \end{equation}
  \item The positive Borel measure $\alpha(s)/\pi$ itself solves the
    symmetric Hamburger moment problem with moments $c'_{2n}=c_n$,
    $c'_{2n+1}=0$.
  \end{enumerate}
\end{theorem}
We recall that the Hamburger moment problem is the problem of reconstructing
a probability measure on the real line from the sequence of its
momenta. The symmetric Hamburger problem is the Hamburger problem in
the particular case where the  momenta of odd parity vanish.

The analytic properties of the Laplace transform can then be
immediately read from the properties of the measure $\alpha$,
since the following corollary holds (Corollary~1 of the
of the paper \cite{cargiorgmaioc})
\begin{corollary}\label{cor:analytic}
$\mathcal C_f(t)$ decays exponentially fast as $t$ goes to infinity if
and only if the coefficients $c_n$ are such that $\alpha'(s)$ exists
and is analytic.
\end{corollary}

Such a result opens up the possibility to study the asymptotic decay
of correlations through the methods developed for the moment problem
(see \cite{Akh,Schmu}), a well--known topic of probability theory. In
particular, necessary and sufficient conditions on the sequence of
coefficients $c_n$ (see a summary in Appendix~A of
\cite{cargiorgmaioc}) have been found to determine whether the measure
$\alpha$ is regular, and therefore to determine whether the decay of
the correlations are exponential or not. Since such conditions usually
require knowing the entire sequence of coefficients in sufficient
detail, this method could however not been implemented yet to obtain
rigorous results.

In any case, as already mentioned, the sequence of the first
coefficients can be calculated numerically, and provide an
approximation for the Laplace transform \eqref{eq:stieltjes_laplace}
from which some indications of the character of the decay of
correlations can be extracted. In the same paper \cite{cargiorgmaioc},
an application is done to the FPU system, which suggests that at low
temperatures the measure $\alpha$ could be singular. Another
application was in the paper \cite{amati}, in which a quartic FPU
model was studied, with a choice of potential parameters such that the
model simulates a glassy solid.

\vskip 2.em
\noindent
\bibliographystyle{plain}
\bibliography{RassegnaBCM26.bib}

@inbook{Gallone2022,
  title = {Hamiltonian Field Theory Close to the Wave Equation: From Fermi-Pasta-Ulam to Water Waves},
  ISBN = {9789811964343},
  ISSN = {2281-5198},
  url = {http://dx.doi.org/10.1007/978-981-19-6434-3_10},
  DOI = {10.1007/978-981-19-6434-3_10},
  booktitle = {Qualitative Properties of Dispersive PDEs},
  publisher = {Springer Nature Singapore},
  author = {Gallone,  Matteo and Ponno,  Antonio},
  year = {2022},
  pages = {205–244}
}

@article{Ono18,
  title = {Double Scaling in the Relaxation Time in the 
$\beta$-Fermi-Pasta-Ulam-Tsingou Model},
  volume = {120},
  ISSN = {1079-7114},
  number = {14},
  journal = {Physical Review Letters},
  publisher = {American Physical Society (APS)},
  author = {Lvov,  Yuri V. and Onorato,  Miguel},
  year = {2018},
  month = Apr 
}

@article{Ono15,
  title = {Route to thermalization in the
           $\alpha$ -{F}ermi–{P}asta–{U}lam system},
  volume = {112},
  ISSN = {1091-6490},
  url = {http://dx.doi.org/10.1073/pnas.1404397112},
  DOI = {10.1073/pnas.1404397112},
  number = {14},
  journal = {Proceedings of the National Academy of Sciences},
  publisher = {Proceedings of the National Academy of Sciences},
  author = {Onorato,  Miguel and Vozella,  Lara and Proment,  Davide and Lvov,  Yuri V.},
  year = {2015},
  month = Mar,
  pages = {4208–4213}
}

@article {GPR21,
    AUTHOR = {Gallone, Matteo and Ponno, Antonio and Rink, Bob},
     TITLE = {Korteweg--de {V}ries and {F}ermi-{P}asta-{U}lam-{T}singou:
              asymptotic integrability of quasi unidirectional waves},
   JOURNAL = {J. Phys. A},
  FJOURNAL = {Journal of Physics. A. Mathematical and Theoretical},
    VOLUME = {54},
      YEAR = {2021},
    NUMBER = {30},
     PAGES = {Paper No. 305701, 29},
      ISSN = {1751-8113,1751-8121},
   MRCLASS = {37K10},
  MRNUMBER = {4294860},
MRREVIEWER = {Pavel\ I.\ Naumkin},
       DOI = {10.1088/1751-8121/ac0a2e},
       URL = {https://doi.org/10.1088/1751-8121/ac0a2e},
}

@incollection{FPU,
AUTHOR = {E. Fermi and J.R. Pasta and S.M. Ulam},
TITLE = {Studies of nonlinear problems},
YEAR = 1965,
booktitle = {Collected works of E. Fermi, vol.2},
publisher ={Chicago University Press},
ADDRESS = {Chicago},
}

@article {Rin01,
    AUTHOR = {Rink, B.},
     TITLE = {Symmetry and resonance in periodic {FPU} chains},
   JOURNAL = {Comm. Math. Phys.},
  FJOURNAL = {Communications in Mathematical Physics},
    VOLUME = {218},
      YEAR = {2001},
    NUMBER = {3},
     PAGES = {665--685},
      ISSN = {0010-3616},
     CODEN = {CMPHAY},
   MRCLASS = {37J40 (70H08 70H33)},
  MRNUMBER = {MR1831098 (2002c:37094)},
MRREVIEWER = {O. Christov},
}

@article {Rin03,
    AUTHOR = {Rink, B.},
     TITLE = {Symmetric invariant manifolds in the {F}ermi-{P}asta-{U}lam
              lattice},
   JOURNAL = {Phys. D},
  FJOURNAL = {Physica D. Nonlinear Phenomena},
    VOLUME = {175},
      YEAR = {2003},
    NUMBER = {1-2},
     PAGES = {31--42},
      ISSN = {0167-2789},
     CODEN = {PDNPDT},
   MRCLASS = {70K42 (37J15 37L60 82C05)},
  MRNUMBER = {MR1957905 (2004b:70048)},
MRREVIEWER = {P. G. Kevrekidis},
}

@article {CN92,
    AUTHOR = {Lochak, P. and Ne\u{\i}shtadt, A. I.},
     TITLE = {Estimates of stability time for nearly integrable systems with
              a quasiconvex {H}amiltonian},
   JOURNAL = {Chaos},
  FJOURNAL = {Chaos. An Interdisciplinary Journal of Nonlinear Science},
    VOLUME = {2},
      YEAR = {1992},
    NUMBER = {4},
     PAGES = {495--499},
      ISSN = {1054-1500,1089-7682},
   MRCLASS = {58F10 (58F05 58F30 70H05)},
  MRNUMBER = {1195881},
MRREVIEWER = {Dmitry\ V.\ Treshch\"{e}v},
       DOI = {10.1063/1.165891},
       URL = {https://doi.org/10.1063/1.165891},
}

@article {ZZ17,
    AUTHOR = {Zhang, Jianlu and Zhang, Ke},
     TITLE = {Improved stability for analytic quasi-convex nearly integrable
              systems and optimal speed of {A}rnold diffusion},
   JOURNAL = {Nonlinearity},
  FJOURNAL = {Nonlinearity},
    VOLUME = {30},
      YEAR = {2017},
    NUMBER = {7},
     PAGES = {2918--2929},
      ISSN = {0951-7715,1361-6544},
   MRCLASS = {37J40 (70H08)},
  MRNUMBER = {3670012},
MRREVIEWER = {Xifeng\ Su},
       DOI = {10.1088/1361-6544/aa72b7},
       URL = {https://doi-org.pros1.lib.unimi.it/10.1088/1361-6544/aa72b7},
}

@article {BM16,
    AUTHOR = {Bambusi, D. and Maspero, A.},
     TITLE = {Birkhoff coordinates for the {T}oda lattice in the limit of
              infinitely many particles with an application to {FPU}},
   JOURNAL = {J. Funct. Anal.},
  FJOURNAL = {Journal of Functional Analysis},
    VOLUME = {270},
      YEAR = {2016},
    NUMBER = {5},
     PAGES = {1818--1887},
      ISSN = {0022-1236,1096-0783},
   MRCLASS = {37K10 (35Q53)},
  MRNUMBER = {3452718},
MRREVIEWER = {Gayrat\ Urazboev},
       DOI = {10.1016/j.jfa.2015.08.003},
       URL = {https://doi.org/10.1016/j.jfa.2015.08.003},
}

@article {GCB,
    AUTHOR = {Guzzo, M. and Chierchia, L. and Benettin, G.},
     TITLE = {The steep {N}ekhoroshev's theorem},
   JOURNAL = {Comm. Math. Phys.},
  FJOURNAL = {Communications in Mathematical Physics},
    VOLUME = {342},
      YEAR = {2016},
    NUMBER = {2},
     PAGES = {569--601},
      ISSN = {0010-3616,1432-0916},
   MRCLASS = {37J25},
  MRNUMBER = {3459160},
MRREVIEWER = {Xifeng\ Su},
       DOI = {10.1007/s00220-015-2555-x},
       URL = {https://doi.org/10.1007/s00220-015-2555-x},
}

@article {BKP0,
    AUTHOR = {Bambusi, Dario and Kappeler, Thomas and Paul, Thierry},
     TITLE = {De {T}oda \`a {K}d{V}},
   JOURNAL = {C. R. Math. Acad. Sci. Paris},
  FJOURNAL = {Comptes Rendus Math\'{e}matique. Acad\'{e}mie des Sciences.
              Paris},
    VOLUME = {347},
      YEAR = {2009},
    NUMBER = {17-18},
     PAGES = {1025--1030},
      ISSN = {1631-073X,1778-3569},
   MRCLASS = {35Q53},
  MRNUMBER = {2554570},
       DOI = {10.1016/j.crma.2009.07.002},
       URL = {https://doi.org/10.1016/j.crma.2009.07.002},
}

@article {BKP1,
    AUTHOR = {Bambusi, D. and Kappeler, T. and Paul, T.},
     TITLE = {From {T}oda to {K}d{V}},
   JOURNAL = {Nonlinearity},
  FJOURNAL = {Nonlinearity},
    VOLUME = {28},
      YEAR = {2015},
    NUMBER = {7},
     PAGES = {2461--2496},
      ISSN = {0951-7715,1361-6544},
   MRCLASS = {37K10 (35Q53 70F45 81R12)},
  MRNUMBER = {3366652},
MRREVIEWER = {Rajkumar\ Roychoudhury},
       DOI = {10.1088/0951-7715/28/7/2461},
       URL = {https://doi.org/10.1088/0951-7715/28/7/2461},
}

@article {Craig,
    AUTHOR = {Craig, Walter and Groves, Mark D.},
     TITLE = {Hamiltonian long-wave approximations to the water-wave
              problem},
   JOURNAL = {Wave Motion},
  FJOURNAL = {Wave Motion. An International Journal Reporting Research on
              Wave Phenomena},
    VOLUME = {19},
      YEAR = {1994},
    NUMBER = {4},
     PAGES = {367--389},
      ISSN = {0165-2125,1878-433X},
   MRCLASS = {76B15 (35Q53 58F05 58F40 76M99)},
  MRNUMBER = {1285131},
MRREVIEWER = {David\ J.\ Kaup},
       DOI = {10.1016/0165-2125(94)90003-5},
       URL = {https://doi.org/10.1016/0165-2125(94)90003-5},
}

@incollection {Kodama,
    AUTHOR = {Hiraoka, Y. and Kodama, Y.},
     TITLE = {Normal form and solitons},
 BOOKTITLE = {Integrability},
    SERIES = {Lecture Notes in Phys.},
    VOLUME = {767},
     PAGES = {175--214},
 PUBLISHER = {Springer, Berlin},
      YEAR = {2009},
      ISBN = {978-3-540-88110-0},
   MRCLASS = {37K40 (35C08 35Q53 37K05 37K10)},
  MRNUMBER = {2867550},
MRREVIEWER = {Dmitry\ E.\ Pelinovsky},
       DOI = {10.1007/978-3-540-88111-7\{_}6}

@article {BP06,
    AUTHOR = {Bambusi, Dario and Ponno, Antonio},
     TITLE = {On metastability in {FPU}},
   JOURNAL = {Comm. Math. Phys.},
  FJOURNAL = {Communications in Mathematical Physics},
    VOLUME = {264},
      YEAR = {2006},
    NUMBER = {2},
     PAGES = {539--561},
      ISSN = {0010-3616,1432-0916},
   MRCLASS = {37K60 (37K55 70K45 70K65 82C05)},
  MRNUMBER = {2215616},
MRREVIEWER = {Karsten\ Matthies},
       DOI = {10.1007/s00220-005-1488-1},
       URL = {https://doi.org/10.1007/s00220-005-1488-1},
}

@article {Bam21,
    AUTHOR = {Bambusi, Dario},
     TITLE = {Hamiltonian studies on counter-propagating water waves},
   JOURNAL = {Water Waves},
  FJOURNAL = {Water Waves. An Interdisciplinary Journal},
    VOLUME = {3},
      YEAR = {2021},
    NUMBER = {1},
     PAGES = {49--83},
      ISSN = {2523-367X,2523-3688},
   MRCLASS = {35Q35 (76B15)},
  MRNUMBER = {4246388},
MRREVIEWER = {Koji\ Ohkitani},
       DOI = {10.1007/s42286-020-00032-y},
       URL = {https://doi.org/10.1007/s42286-020-00032-y},
}

@article {Kuk96,
    AUTHOR = {Kuksin, Sergei B.},
     TITLE = {Growth and oscillations of solutions of nonlinear
              {S}chr\"{o}dinger equation},
   JOURNAL = {Comm. Math. Phys.},
  FJOURNAL = {Communications in Mathematical Physics},
    VOLUME = {178},
      YEAR = {1996},
    NUMBER = {2},
     PAGES = {265--280},
      ISSN = {0010-3616,1432-0916},
   MRCLASS = {35Q55 (35B05 58F39)},
  MRNUMBER = {1389904},
MRREVIEWER = {J\"{u}rgen\ P\"{o}schel},
       URL = {http://projecteuclid.org/euclid.cmp/1104286651},
}

@article {DH23,
    AUTHOR = {Deng, Yu and Hani, Zaher},
     TITLE = {Full derivation of the wave kinetic equation},
   JOURNAL = {Invent. Math.},
  FJOURNAL = {Inventiones Mathematicae},
    VOLUME = {233},
      YEAR = {2023},
    NUMBER = {2},
     PAGES = {543--724},
      ISSN = {0020-9910,1432-1297},
   MRCLASS = {35Q55},
  MRNUMBER = {4607721},
       DOI = {10.1007/s00222-023-01189-2},
       URL = {https://doi.org/10.1007/s00222-023-01189-2},
}

@misc{DHM24,
      title={Long time derivation of the {B}oltzmann equation from hard sphere dynamics.\ arXiv:2408.07818}, 
      author={Yu Deng and Zaher Hani and Xiao Ma},
      year={2024},
      eprint={2408.07818},
      archivePrefix={arXiv},
      primaryClass={math.AP},
      url={https://arxiv.org/abs/2408.07818}, 
}

@misc{OnoHani,
      title={Validity condition of normal form transformation for the $\beta$-FPUT system. arxiv:2510.04831}, 
      author={Boyang Wu and Miguel Onorato and Zaher Hani and Yulin Pan},
      year={2025},
      eprint={2510.04831},
      archivePrefix={arXiv},
      primaryClass={math-ph},
      url={https://arxiv.org/abs/2510.04831}, 
}

@misc{Wu,
      title={Rigorous Derivation of the {W}ave {K}inetic {E}quation for $\beta$-FPUT System. arxiv:2506.02948}, 
      author={Wu, B},
      year={2025},
      eprint={2506.02948},
      archivePrefix={arXiv},
}

@article {GPRuffo,
    AUTHOR = {Gallone, Matteo and Marian, Matteo and Ponno, Antonio and
              Ruffo, Stefano},
     TITLE = {Burgers turbulence in the {F}ermi-{P}asta-{U}lam-{T}singou
              chain},
   JOURNAL = {Phys. Rev. Lett.},
  FJOURNAL = {Physical Review Letters},
    VOLUME = {129},
      YEAR = {2022},
    NUMBER = {11},
     PAGES = {Paper No. 114101, 6},
      ISSN = {0031-9007,1079-7114},
   MRCLASS = {82C20},
  MRNUMBER = {4490374},
       DOI = {10.1103/physrevlett.129.114101},
       URL = {https://doi.org/10.1103/physrevlett.129.114101},
}

@article {BKP2,
    AUTHOR = {Bambusi, D. and Kappeler, T. and Paul, T.},
     TITLE = {Dynamics of periodic {T}oda chains with a large number of
              particles},
   JOURNAL = {J. Differential Equations},
  FJOURNAL = {Journal of Differential Equations},
    VOLUME = {258},
      YEAR = {2015},
    NUMBER = {12},
     PAGES = {4209--4274},
      ISSN = {0022-0396,1090-2732},
   MRCLASS = {37K10 (35B10 35Q53)},
  MRNUMBER = {3327553},
MRREVIEWER = {Am\'{\i}lcar\ Branquinho},
       DOI = {10.1016/j.jde.2015.01.031},
       URL = {https://doi.org/10.1016/j.jde.2015.01.031},
}

@article {HK08a,
    AUTHOR = {Henrici, Andreas and Kappeler, Thomas},
     TITLE = {Results on normal forms for {FPU} chains},
   JOURNAL = {Comm. Math. Phys.},
  FJOURNAL = {Communications in Mathematical Physics},
    VOLUME = {278},
      YEAR = {2008},
    NUMBER = {1},
     PAGES = {145--177},
      ISSN = {0010-3616,1432-0916},
   MRCLASS = {37J40 (37K60 70F10 70H06 70K45 81R12 82B05)},
  MRNUMBER = {2367202},
MRREVIEWER = {Jens\ D. M. Rademacher},
       DOI = {10.1007/s00220-007-0387-z},
       URL = {https://doi.org/10.1007/s00220-007-0387-z},
}

@article {HK09a,
    AUTHOR = {Henrici, Andreas and Kappeler, Thomas},
     TITLE = {Nekhoroshev theorem for the periodic {T}oda lattice},
   JOURNAL = {Chaos},
  FJOURNAL = {Chaos. An Interdisciplinary Journal of Nonlinear Science},
    VOLUME = {19},
      YEAR = {2009},
    NUMBER = {3},
     PAGES = {033120, 13},
      ISSN = {1054-1500,1089-7682},
   MRCLASS = {37J35 (30F30 34A33 70H06)},
  MRNUMBER = {2573925},
MRREVIEWER = {Mikhail\ B.\ Sevryuk},
       DOI = {10.1063/1.3196783},
       URL = {https://doi.org/10.1063/1.3196783},
}

@article {HK09,
    AUTHOR = {Henrici, Andreas and Kappeler, Thomas},
     TITLE = {Resonant normal form for even periodic {FPU} chains},
   JOURNAL = {J. Eur. Math. Soc. (JEMS)},
  FJOURNAL = {Journal of the European Mathematical Society (JEMS)},
    VOLUME = {11},
      YEAR = {2009},
    NUMBER = {5},
     PAGES = {1025--1056},
      ISSN = {1435-9855,1435-9863},
   MRCLASS = {37J15 (37J35 70H06 70H08)},
  MRNUMBER = {2538499},
MRREVIEWER = {Jens\ D. M. Rademacher},
       DOI = {10.4171/JEMS/174},
       URL = {https://doi.org/10.4171/JEMS/174},
}

@article {HK08,
    AUTHOR = {Henrici, Andreas and Kappeler, Thomas},
     TITLE = {Global {B}irkhoff coordinates for the periodic {T}oda lattice},
   JOURNAL = {Nonlinearity},
  FJOURNAL = {Nonlinearity},
    VOLUME = {21},
      YEAR = {2008},
    NUMBER = {12},
     PAGES = {2731--2758},
      ISSN = {0951-7715,1361-6544},
   MRCLASS = {37J35 (37K60 70H06)},
  MRNUMBER = {2461037},
       DOI = {10.1088/0951-7715/21/12/001},
       URL = {https://doi.org/10.1088/0951-7715/21/12/001},
}

@article{douglas,
author ="Douglas, Jack F. and Yuan, Qi-Lu and Zhang, Jiarui and Zhang, Hao and Xu, Wen-Sheng",
title  ="A dynamical system approach to relaxation in glass-forming liquids",
journal  ="Soft Matter",
year  ="2024",
volume  ="20",
issue  ="46",
pages  ="9140-9160",
publisher  ="The Royal Society of Chemistry",
doi  ="10.1039/D4SM00976B",
url  ="http://dx.doi.org/10.1039/D4SM00976B",
}

@article{bamcarmaio,
author ="Maiocchi, A. and Bambusi, D. and Carati, A.",
title  ="An Averaging Theorem for {FPU} in the Thermodynamic Limit",
journal ="J. Stat. Phys.",
year  ="2014",
volume  ="155",
issue  ="",
pages  ="300-322",
publisher  ="",
doi  ="10.1007/s10955-014-0958-2",
url  =""
}

@article{Carati,
author ="Carati, A.",
title  ="An averaging theorem for Hamiltonian dynamical systems in the
thermodynamic limit",
journal ="J. Stat. Phys.",
year  ="2007",
volume  ="128",
issue  ="",
pages  ="1057-1077",
publisher  ="",
doi  ="10.1007/s10955-007-9332-y",
url  =""
}

@article{Maiocchi,
author ="Maiocchi, A.M.",
title  ="Freezing of the Optical-Branch Energy in a Diatomic {FPU} Chain",
journal ="Comm. Math. Phys.",
year  ="2019",
volume  ="372",
issue  ="",
pages  ="91-117",
publisher  ="",
doi  ="10.1007/s00220-019-03381-z",
url  =""
}

@article{cargiorgmaioc,
author ="Maiocchi, A.M. and Carati, A. and Giorgilli, A.",
title  ="A Series Expansion for the Time Autocorrelation of Dynamical Variables",
journal ="J. Stat. Phys.",
year  ="2012",
volume  ="148",
issue  ="",
pages  ="1054-1071",
publisher  ="",
doi  ="10.1007/s10955-012-0575-x",
url  =""
}

@article{amati,
author ="Amati, G. and Meyer, H. and Schilling, T.",
title  ="Memory Effects in the {F}ermi–{P}asta–{U}lam Model",
journal ="J. Stat. Phys.",
year  ="2019",
volume  ="174",
issue  ="",
pages  ="219-257",
publisher  ="",
doi  ="10.1007/s10955-018-2207-6",
url  =""
}

@article{benchripon,
author ="Benettin, G. and Christodoulidi, H. and Ponno, A.",
title  ="The {F}ermi–{P}asta–{U}lam and Its Underlying Integrable Dynamics",
journal ="J. Stat. Phys.",
year  ="2013",
volume  ="152",
issue  ="",
pages  ="195-212",
publisher  ="",
doi  ="10.1007/s10955-013-0760-6",
url  =""
}

@article{GMMP,
author ="Grava, T. and Maspero, A. and Mazzuca, G. and Ponno, A.",
title  ="Adiabatic invatiants for the {FPUT} and {T}oda chain in the thermodynamic limit",
journal ="Commun. Math. Phys.",
year  ="2020",
volume  ="380",
issue  ="",
pages  ="811-851",
publisher  ="",
doi  ="10.1007/s00220-020-03866-2",
url  =""
}

@article{BK,
author ="von Kármán, T. and Born, M.",
title  ="{\"U}ber {S}chwingungen in {R}aumgittern",
journal ="Physikalische Zeitschrift",
year  ="1913",
volume  ="13",
issue  ="",
pages  ="297-309",
publisher  ="",
doi  ="",
url  =""
}

@article{GGMV,
author ="Galgani, L. and Giorgilli, A. and Martinoli, A. and Vanzini, S.",
title  ="On the problem of energy equipartition for large systems of the
Fermi-Pasta-Ulam type: analytical and numerical estimates,",
journal ="Physica D",
year  ="1992",
volume  ="59",
issue  ="",
pages  ="334--348",
publisher  ="",
doi  ="10.1016/0167-2789(92)90074-W",
url  =""
}

@book{Akh,
author ="Akhiezer, N.I.",
title  ="The classical moment problem",
year  ="1965",
pages  ="",
publisher  ="Oliver and Boyd",
doi  ="",
url  ="",
address = "London and Edinburgh"
}

@book{Schmu,
author ="Schm{\"u}dgen, K.",
title  ="The moment problem",
series = "Graduate texts in mathematics",
year  ="2017",
volume  ="277",
publisher  ="Springer",
doi  ="10.1007/978-3-319-64546-9",
url  ="",
address = "Cham"
}

\end{document}